\documentclass{article}

\usepackage{amssymb}
\usepackage{amsmath}
\usepackage{graphicx}
\usepackage{caption}
\usepackage{natbib}
\usepackage{rotating}
\usepackage{mathtools}
\usepackage{multirow}
\usepackage{tabularx}
\usepackage{scrextend}
\usepackage{hyperref}
\usepackage{authblk}

\newcolumntype{L}[1]{>{\raggedright\let\newline\\\arraybackslash\hspace{0pt}}m{#1}}
\newcolumntype{C}[1]{>{\centering\let\newline\\\arraybackslash\hspace{0pt}}m{#1}}
\newcolumntype{R}[1]{>{\raggedleft\let\newline\\\arraybackslash\hspace{0pt}}m{#1}}

\providecommand{\keywords}[1]{\textbf{\textit{Keywords:}} #1}

	\title{Multiple imputation in data that grow over time: A comparison of three strategies} 

	\author{X.M. Kavelaars$^{1}$\thanks{Electronic address: \texttt{x.m.kavelaars@tilburguniversity.edu}; Corresponding author} , S. Van Buuren$^{2,3}$, and J.R. Van Ginkel$^4$} 

	\date{\footnotesize $^1$  Department of Methodology and Statistics, Tilburg University, The Netherlands\\
	$^2$ \footnotesize Department of Methodology and Statistics, Tilburg University, The Netherlands\\
	$^3$ \footnotesize Department of Statistics, TNO Quality of Life, Leiden, The Netherlands\\
	$^4$ \footnotesize Department of Psychology, Leiden University, The Netherlands}

\begin{document}

	\maketitle 
	
		\begin{abstract}Multiple imputation is a highly recommended technique to deal with missing data, but the application to longitudinal datasets can be done in multiple ways. When a new wave of longitudinal data arrives, we can treat the combined data of multiple waves as a new missing data problem and overwrite existing imputations with new values (re-imputation). Alternatively, we may keep the existing imputations, and impute only the new data. We may do either a full multiple imputation (nested) or a single imputation (appended) on the new data per imputed set. This study compares these three strategies by means of simulation. All techniques resulted in valid inference under a monotone missingness pattern. A non-monotone missingness pattern led to biased and non-confidence valid regression coefficients after nested and appended imputation, depending on the correlation structure of the data. Correlations within timepoints must be stronger than correlations between timepoints to obtain valid inference. In an empirical example, the three strategies performed similarly. We conclude that appended imputation is especially beneficial in longitudinal datasets that suffer from dropout.
	\end{abstract}

	\section{Introduction}\label{sec:introduction}
	
	Incomplete data are nearly inevitable in medical studies and require careful attention, since missing data can seriously affect the validity of statistical inference when not handled properly \citep{rubin1987}. A commonly used technique to obtain valid inference with incomplete datasets is multiple imputation: A procedure that replaces each missing value by  several plausible values thereby creating multiple completed datasets. The imputation procedure is followed by an analysis procedure in which each dataset is analyzed separately before these individual results are pooled into a single estimate. The pooling of multiple parameter estimates is useful to reflect an additional source of uncertainty that imputed values bring to the analysis: Since we do not know how well imputed values correspond to the unobserved values, differences between imputations are used to quantify and incorporate this uncertainty in statistical inference \citep{rubin1987}.
	
	When the imputation and analysis procedures have been completed, new data may be added to the existing dataset. Such multistage data are common, with longitudinal (new variables) and sequential (new cases) designs as well-known examples. New data raise the question what to do with the existing imputations. We can overwrite the existing imputations with new values or we can keep the existing imputations and impute the new data only \citep{rubin2003, harel2003a, mcginniss2016}. 
	
	\section{Problem illustration}\label{sec:problem}
	We illustrate these methods and their operational aspects with the Project on Preterm and Small for Gestational Age Infants (POPS) \citep{verloove1986,veen1991}. The POPS study included about 94\% of children born in 1983 in The Netherlands with a birth weight below $1500$ grams and/or a gestational age below $32$ weeks ($n=1338$). POPS has multiple waves since these children have been followed at various ages (e.g. 1, 5, 10, 14, 19 years) to evaluate physical, cognitive and psychosocial outcomes. Like many longitudinal studies, POPS suffers from dropout that should not be ignored: Of the $n=959$ surviving participants, $n=596$ completed participation at age 19 and those who dropped out differed systematically from full responders on multiple outcomes of interest \citep{hille2005}. Each wave of data has missing values that we aim to treat with multiple imputation.
		
	Overwriting existing imputations with new values is called re-imputation and treats the combined data of multiple waves as a new missing data problem. This strategy merges the incomplete data of previous waves with new variables and performs a multiple imputation on these combined data. Each new wave of data results in a new set of $m$ imputed datasets. Updating existing imputations with new data has consequences for the analysis of the old data. Repeating previous analyses with newly imputed data may result in different parameter estimates or conclusions. We demonstrate this with an adapted example from van Dommelen et al. \citep{dommelen2014}. The researchers used the POPS data to predict the effect of early catch-up growth (developing towards the median of the growth charts in the first year of life) on health and well-being in young adulthood from the POPS data. Multiple imputation of the data available at age 14 showed that catch-up growth in weight did not predict length and weight at age 14. We observed the opposite when we ran the same analysis with data from age 19 included in the imputation: Catch-up growth did predict length and weight at age 14\footnote[1]{$b_{\text{length }14}= 0.25$, $95\%$ CI $(-0.10-0.60)$, $b_{\text{weight }14}= 0.21$, $95\%$ CI $(-0.04-0.47)$, $b_{\text{length }19}= 0.32$, $95\%$ CI $(0.01-0.62)$, $b_{\text{weight} 19}= 0.33$, $95\%$ CI $(0.09-0.56)$}. 
	Re-imputation potentially lacks replicability and contrasting conclusions as these raise the question which of the analyses should be trusted.
		
	Keeping the existing imputations and imputing only the new data is known as nested imputation \citep{rubin2003}. This strategy combines the completed datasets from previous waves with the new incomplete data and performs a multiple imputation on each of these combined datasets \citep{rubin2003,harel2003a,harel2003b}. The data are imputed $m_i$ times in wave $i$ resulting in a total of $\prod_{i=1} m_i$ nested datasets. Nested imputation preserves completed datasets from previous waves, which is useful since all existing analyses remain unchanged. The strategy can be computationally challenging however, since the number of imputed datasets increases exponentially with each additional wave of data. These datasets require specific pooling rules to take the nested structure into account and the complexity of these rules increases sharply when the number of stages goes up \citep{shen2000, rubin2003, mcginniss2016}. Nested imputation would be highly inconvenient for datasets with several waves.
		
	As a potentially more convenient alternative, we propose to append imputations to the existing ones and do a single imputation on the new data of each of $m_1$ completed datasets. Appended imputation is a special case of nested imputation with $m_i=1$ for $i>1$. The method has several attractive features. Whereas re-imputation includes future information to overwrite existing imputations, the successive structure of appended imputation preserves existing imputations and allows for replicable conclusions. Appended imputation is computationally convenient, since the resulting $m_1$ imputations can be pooled with a procedure for non-nested datasets. 
		
	While appended imputation has operational advantages over re-imputation and nested imputation, differences between strategies may impact on the validity and the efficiency of statistical inference with the completed data. To be more specific, it is unclear whether the initial multiple imputation of appended imputation can account for sufficient imputation uncertainty to obtain unbiased and confidence-valid estimates. Moreover, the difference between imputing the entire dataset (re-imputation) or new data only (nested imputation) in multistage data might influence the validity of statistical inference as well. The current study aims to evaluate the statistical properties of the three introduced strategies for creating imputations in a multistage context. Section \ref{sec:assumptions} discusses potential effects of updating or preserving existing imputations on validity. In Section \ref{sec:simulation}, the statistical performance of the three strategies are evaluated by means of simulation. We apply the three strategies to the POPS dataset in Section \ref{sec:application} and conclude the paper with a discussion in Section \ref{sec:discussion}.
	
	\section{Assumptions}\label{sec:assumptions}
	
	\subsection{Congeniality}
	When the statistical model of interest (i.e. the analysis model) describes relations between variables, unobserved values should be imputed in such a way that the completed data properly reflect these relationships. This condition has been met if the variables in the imputation model correspond to the variables in the analysis model: An assumption that we call congeniality between the imputation and the analysis \citep{meng1994}. Any discrepancy between the variables in the imputation model and the analysis model results in an uncongenial imputation model.
	
	Uncongeniality can be problematic for the validity of statistical inference, in particular when the imputation models contains fewer variables than the analysis model. Ignoring variables in the imputation model assumes there is no relationship between the incomplete and the omitted variable, resulting in subjects with (falsely) unrelated imputations \citep{meng1994}. Mixing these subjects with (correctly) correlated observed values results in systematical underestimation of the strength of this relationship and undercoverage of nominal confidence intervals of relationship parameters \citep{meng1994,zhu2015,Xie2015,daniels2014,gelman2001}. In contrast, validity is generally not threatened by uncongeniality if the imputation model contains more variables than the analysis model. The information from extra variables results in so-called superefficient parameter estimates, that have smaller variances than can be obtained with a congenial imputation model \citep{rubin1996,Xie2015}. 
	
	The three imputation strategies include new variables in the imputation models in different ways and consequentially deal with congeniality differently. Re-imputation updates existing imputations with information obtained from the new variables and can result in congenial imputation models for analysis models that relate old and new variables. If the analysis model is limited to variables from previous waves while the imputation model includes new variables, the updated imputations include extra information and may result in sharper inferences than the analysis using the existing imputations (which is illustrated by the example in Section \ref{sec:problem}). 
	
	Since nested and appended imputation ignore any information available from later waves, existing imputations were created under the assumption that the old variables are unrelated to the new variables. Although the imputation model of the existing imputations and any analysis model relating old and new values are thus uncongenial, estimates of nonzero relationships are not necessarily incorrect. Whether these parameters are estimated accurately, depends on the combination of missing values as reflected by the missingness pattern. Estimates may be correct under monotone missingness, which is characterized by the following two situations: 1.) Incomplete cases have missing data in the new variables only; 2.) Incomplete cases have missing data on old variables and \textit{all} new variables. In contrast, relationships will be underestimated when missingness is nonmonontone, such that incomplete cases have missing data in the old variables while the new variables have been observed. Newly observed values might provide essential information about the unobserved values in the old data, if these old and new variables are related. The existing imputations could not take this information into account since it was not in their imputation model. Monotone missingness does not suffer from biased relationships: There are no newly observed values to provide information about the existing imputations and their relations with new variables.

	\subsection{Imputation uncertainty}
	Proper quantification of uncertainty requires reflection of imputation uncertainty: The variance of parameter estimates must correctly reveal the extra uncertainty of adopting imputations as data \citep{rubin1987}. The pooling procedure therefore accommodates differences between imputed datasets, such that parameter estimates are confidence-valid. Whereas re-imputation uses regular pooling rules for multiple imputation, nested multiple imputation requires specific pooling rules that respect the nested data structure \citep{rubin2003,harel2003a,shen2000}. These pooling rules for two-level nested datasets are more complex than the standard pooling rules for non-nested data, as shown in Table \ref{tab:pool} \citep{shen2000,mcginniss2016}.

	Datasets completed with appended imputation have nests of $m_i=1$, such that the pooling procedure for nested imputation simplify to the regular pooling rules \citep{rubin2003}. However, single imputation in general underestimates variance \citep{rubin1987}, and it is unclear whether the multiple imputation structure of the first stage can account for sufficient imputation uncertainty to justify single imputation in later waves. A full multiple imputation at each later stage (i.e. nested imputation) may be necessary to obtain confidence-valid results.
	
	\subsection{Summary}
	To conclude, nested and appended imputation in general make more restrictive assumptions about the missingness pattern of incomplete data than re-imputation, and appended imputation may underestimate variance. In the current simulation study, we investigated these assumptions and their consequences for statistical inference, resulting in the following research questions:
	\begin{enumerate}
		\item In which situations could possible uncongeniality between the imputation and complete-data models created under nested and appended imputation affect the validity of statistical inferences?
		\item Would the three imputation strategies perform similarly under a monotone missing data mechanism?
		\item Does appended imputation result in confidence-valid parameter estimates?
	\end{enumerate}
	
	\section{Simulation study}\label{sec:simulation}	
	\subsection{Simulation setup}
	We evaluated the perfomances of re-imputation, nested imputation and appended imputation in a two-stage setup, resembling a longitudinal design with two waves. The $t_1$ data consisted of one completely observed ($x_1$) and one incomplete ($y_1$) covariate. The $t_2$ data had two incomplete variables ($x_2$ and $y_2$). To investigate the consequences of potential uncongeniality, we manipulated the correlation structure and the missingness pattern of the data. 
	
	Since the relation between old and new data is crucial for valid inference, we distinguished correlations between variables within waves ($\rho_{\text{within}} = \rho_{x_1y_1}, \rho_{x_2y_2}$) from correlations between variables from different waves ($\rho_{\text{between}}=\rho_{x_1x_2},\rho_{x_1y_2},\rho_{y_1x_2},\rho_{y_1y_2}$).
	We specified 16 correlation structures ($\rho_{\text{within}} = 0.1, 0.3, 0.5, \text{ or } 0.7; \rho_{\text{between}} = 0.1, 0.3, 0.5, \text{ or } 0.7$), as presented in Table \ref{tab:design}.
	
	For each of these correlation structures, we drew $2000$ samples of $n=425$ from the multivariate standard normal distribution. These data were made incomplete under monotone and non-monotone missingness (see Table \ref{tab:mp}) by randomly deleting values in accordance with the missingness pattern. Each combination of missing values had the same probability and every sample had a total of 20\% missing values. 
	
	The samples were completed using re-imputation, nested imputation and appended imputation. Under re-imputation, datasets with incomplete $t_1$ and $t_2$ data were imputed $m=5$ times. Under nested and appended imputation, we imputed each $t_1$ dataset $m_1=5$ times and added incomplete $t_2$ data to each completed $t_1$ dataset. We imputed these partially completed datasets $m_2=5$ (nested imputation) and $m_2=1$ (appended imputation) times, resulting in a total of 5 datasets after re-imputation and appended imputation, and 25 datasets after nested imputation.
	
	\begin{table}
		\caption{Different correlation structures used in the simulation study}
		\centering
		\label{tab:design}
		\begin{minipage}{\textwidth}
		\begin{tabular}{cll p{0.05\textwidth} llll}
			\hline
			&\multicolumn{2}{l}{$\rho_{\text{within}}$} && \multicolumn{4}{l}{$\rho_{\text{between}}$}\\
			Scenario & $\rho_{x_1y_1}$ & $\rho_{x_2y_2}$ & & $\rho_{x_1x_2}$ & $\rho_{y_1y_2}$ & $\rho_{y_1x_2}$ & $\rho_{x_1y_2}$ \\ \hline
			1. & $0.1$ & $0.1$ && $0.1$ & $0.1$ & $0.1$ & $0.1$\\
			2. & 0.1 & 0.1 && 0.3 & 0.3 & 0.3 & 0.3\\
			3. & 0.1 & 0.1 && 0.5 & 0.5 & 0.5 & 0.5\\
			4. & 0.1 & 0.1 && 0.7 & 0.7 & 0.7 & 0.7\\
			5. & 0.3 & 0.3 && 0.1 & 0.1 & 0.1 & 0.1\\
			6. & 0.3 & 0.3 && 0.3 & 0.3 & 0.3 & 0.3\\
			7. & 0.3 & 0.3 && 0.5 & 0.5 & 0.5 & 0.5\\
			8. & 0.3 & 0.3 && 0.7 & 0.7 & 0.7 & 0.7\\
			9. & 0.5 & 0.5 && 0.1 & 0.1 & 0.1 & 0.1\\
			10. & 0.5 & 0.5 && 0.3 & 0.3 & 0.3 & 0.3\\
			11. & 0.5 & 0.5 && 0.5 & 0.5 & 0.5 & 0.5\\
			12. & 0.5 & 0.5 && 0.7 & 0.7 & 0.7 & 0.7\\
			13. & 0.7 & 0.7 && 0.1 & 0.1 & 0.1 & 0.1\\
			14. & 0.7 & 0.7 && 0.3 & 0.3 & 0.3 & 0.3\\
			15. & 0.7 & 0.7 && 0.5 & 0.5 & 0.5 & 0.5\\
			16. & 0.7 & 0.7 && 0.7 & 0.7 & 0.66\footnote{\label{1sttablefoot} We slightly lowered $\rho_{y_1x_2}$ and $\rho_{x_1y2}$ to ensure positive definiteness of the covariance matrix.}  & 0.66\footref{1sttablefoot} \\ \hline
		\end{tabular}
	\end{minipage}	
	\end{table}

	\begin{table}[ht]
		\centering
		\caption{Possible combinations of complete (1) and incomplete (0) variables for monotone and non-monotone missing data patterns.} 
		\label{tab:mp}
		\begin{tabular}{ccccccccccc}
			\multicolumn{5}{c}{Monotone}&&
			\multicolumn{5}{c}{Non-monotone}\\ \hline
			& \multicolumn{2}{c}{$t_1$} &  \multicolumn{2}{c}{$t_2$ data} & &
			& \multicolumn{2}{c}{$t_1$} &  \multicolumn{2}{c}{$t_2$ data} \\
			& $x_1$ & $y_1$ & $x_2$ & $y_2$ &&& $x_1$  & $y_1$ & $x_2$ & $y_2$ \\ \hline
			1.&1&1&1&1&&1.&1&1&1&1\\
			2.&1&1&1&0&&2.&1&1&1&0\\
			3.&1&1&0&1&&3.&1&1&0&1\\
			4.&1&1&0&0&&4.&1&1&0&0\\
			5.&1&0&0&0&&5.&1&0&0&0\\
			&&&&&&6.&1&0&1&1\\
			&&&&&&7.&1&0&1&0\\
			&&&&&&8.&1&0&0&1\\
		\end{tabular}
	\end{table}
	
	After imputation, a linear regression model was fitted on each completed $t_2$ dataset:
	\begin{equation*}
	\hat{y}_2 = b_0 + b_{x_1} x_1 + b_{y_1} y_1 + b_{x_2} x_2 
	\end{equation*}
	
	Results were pooled using the method-specific pooling rules presented in Table \ref{tab:pool}. Quantities of interest were variable means and regression coefficients of the fitted model. True values were population means $\boldsymbol{\mu}$ and population regression coefficients derived from the correlation structure of the data.	
	
	\begin{table}[!htbp]
		\caption{Pooling rules for multiple imputation with independent datasets (re-imputation and appended imputation) \citep{rubin1987} and two-level nested datasets (nested imputation) \citep{shen2000, rubin2003}.}
		\label{tab:pool}
		\centering
		\setlength{\extrarowheight}{20pt}
			\begin{tabularx}{\textwidth}{>{\hsize=0.12\hsize}X>{\hsize=0.36\hsize}X>{\hsize=0.52\hsize}X}
				\hline
				Parameter & Independent datasets & Nested datasets  \\
				\hline
				Mean \newline estimate
				&
				$\!\begin{aligned}[t]
				\bar{Q}=\frac{1}{m_i}\sum^{m_i}_{k=1} \hat{Q}^{(k)}
				\end{aligned}$ 
				&
				$\!\begin{aligned}[t]
				\bar{Q}=\frac{1}{m_1m_2}\sum^{m_1}_{k=1}\sum^{m_2}_{l=1} \hat{Q}^{(k,l)}
				\end{aligned}$ 
				\\ 
				
				Nest mean &
				- &
				$\!\begin{aligned}[t]
				\bar{Q}_{k}=\frac{1}{m_2}\sum^{m_2}_{l=1}\hat{Q}^{(k,l)}
				\end{aligned}$ 
				\\ 
				
				Sampling \newline variance 
				&
				$\!\begin{aligned}[t]
				\bar{U}=\frac{1}{m_i}\sum^{m_i}_{k=1}{\bar U}_{k}. 
				\end{aligned}$ 
				&
				$\!\begin{aligned}[t]
				\bar{U}=\frac{1}{m_1m_2}\sum^{m_1}_{k=1}\sum^{m_2}_{l=1}\bar U^{(k, l)}
				\end{aligned}$ 
				\\ 
				
				Within \newline variance 
				& - 
				&
				$\!\begin{aligned}[t]
				W = \frac{1}{m_1 (m_2-1)}\sum^{m_1}_{k=1}\sum^{m_2}_{l=1}(\hat{Q}^{(k, l)}-\bar{Q}_{k})^2
				\end{aligned}$ 
				\\ 
				
				Between \newline variance & 
				$\!\begin{aligned}[t]
				B=\frac{1}{m_i-1}\sum^{m_i}_{k=1}(\hat{ Q}_k-\bar{Q})^2.
				\end{aligned}$ 
				&
				$\!\begin{aligned}[]
				B = \frac{m_2}{m_1-1}\sum^{m_1}_{k=1}(\bar{Q}_{k}-\bar{Q})^2
				\end{aligned}$ 
				\\ 
				
				Total \newline variance & 
				$\!\begin{aligned}[t]
				T=\bar{U}+(1+\frac{1}{m_i})B. 
				\end{aligned}$ 
				&
				$\!\begin{aligned}[t]
				&T=\bar{U}+\frac{1}{m_2}(1+\frac{1}{m_1})B +(1-\frac{1}{m_2})W
				\end{aligned}$
				\\ 
				
				Degrees of \newline freedom & 
				$\!\begin{aligned}[t]
				&\nu^{-1} =\frac{1}{m_i-1}\left[\frac{(1+\frac{1}{m_i})B}{T}\right]^2%
				\end{aligned}$
				&
				$\!\begin{aligned}[t]
				\nu^{-1} =&\frac{1}{m_1-1}\left[\frac{\frac{1}{m_2}(1+\frac{1}{m_1})B}{T}\right]^2+\\
				&\frac{1}{m_1(m_2-1)}\left[\frac{(1-\frac{1}{m_2})W}{T}\right]^2
				\end{aligned}$ 
				\\ 
				\hline
				\multicolumn{3}{l}{\raggedright \textit{Note.} $\hat{Q}^{(k,l)}$ and $\bar U^{(k, l)}$ represent the complete data estimate and sampling variance }\\[-20pt]
				\multicolumn{3}{l}{\raggedright for dataset $l$ in nest $k$ respectively.}\\[-20pt]
		\end{tabularx}
	\end{table}
	
	We operationalized performance as validity and efficiency. Imputations were considered valid if pooled parameter estimates were unbiased and the coverage of their confidence intervals was at least the nominal 95\%, adhering to the criteria for proper imputation \citep{rubin1987}. Relative efficiency referred to the width of the 95\% confidence interval compared to the other imputation strategies.
	
	We performed the simulation study in \texttt{R} \citep{RCT2016}. Data amputations and imputations were performed with \texttt {mice}	\citep{buuren2011} using Bayesian linear regression.

	\subsection{Results}
	\subsubsection{Validity}	
	While all analyses following re-imputation resulted in valid estimates (bias: $\leq0.02$,  coverage: $93.1 -96.5\%$), the validity of nested and appended imputation depended on the missingness pattern and the correlation structure of the data. Under monotone missingness, variable means and regression coefficients were valid (bias: $\leq|0.01|$,  coverage: $93.2-96.8\%$) and non-monotone missingness resulted in proper variable mean estimation (bias: $\leq0.01$, coverage: $93.3-97.2\%$). The validity of regression coefficients under non-monotone missingness depended on the correlation structure in the data. 
		
\begin{figure*}[!h]
	\centering
	\includegraphics[width=\textwidth,keepaspectratio]{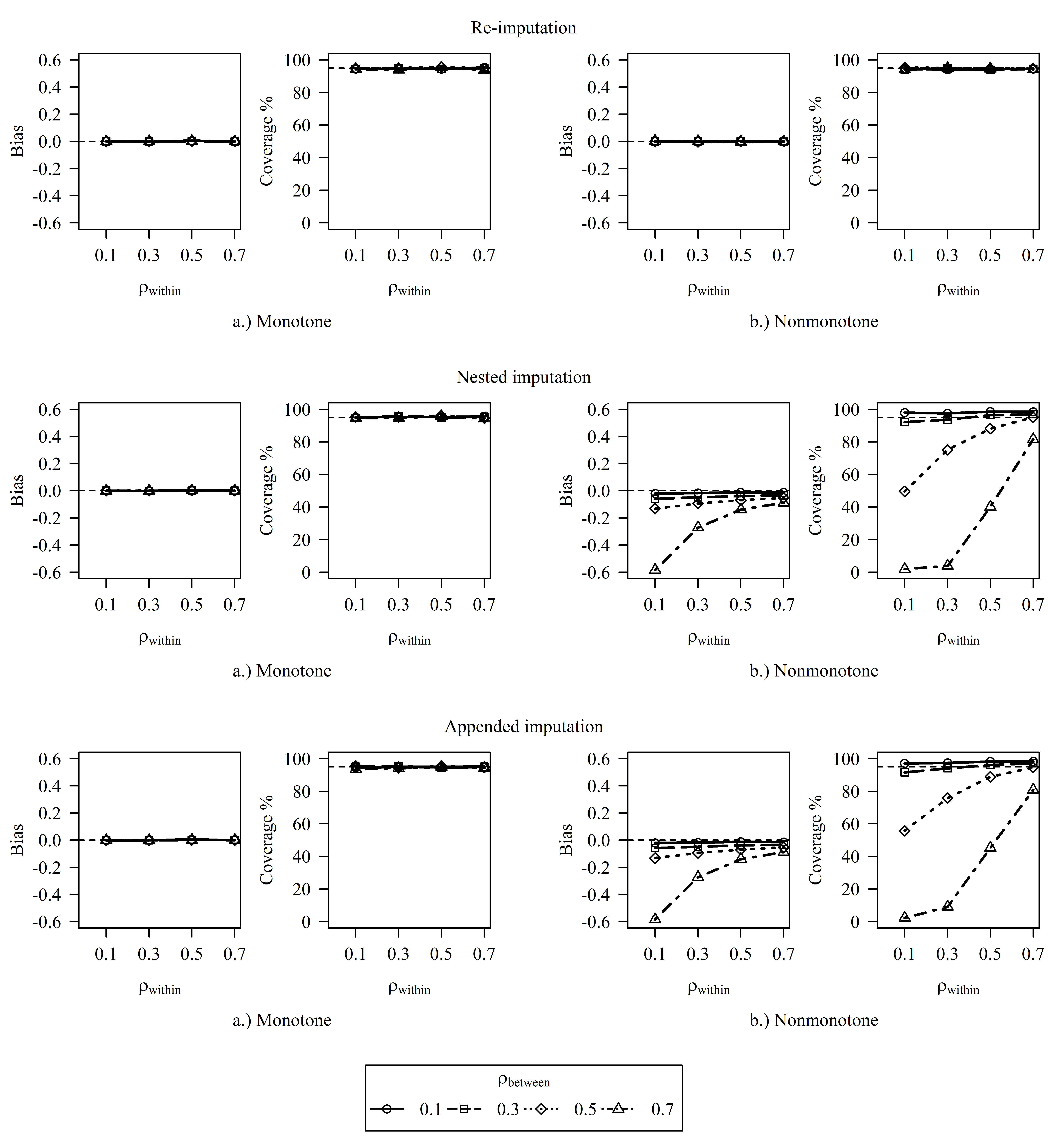}
	\captionsetup{type=figure}
	\caption{Observed patterns of bias and coverage of regression coefficients by correlation structure ($\rho_{\text{within}}=$ correlation within timepoints; $\rho_{\text{between}}=$ correlation between timepoints) after re-imputation, nested imputation and appended imputation under a (a) monotone or (b) non-monotone missingness pattern.}
	\label{fig:result}
\end{figure*}

	When correlations within timepoints were lower than correlations between timepoints, regression coefficient estimates were clearly biased (bias: $0.06-0.59$) and undercovered (coverage: $1.9-92.1\%$). In contrast, when correlations of variables within waves dominated correlations of variables between waves, bias of regression coefficients was small (bias: $0.01-0.05$) and their coverage exceeded the nominal level (coverage: $95.1-98.5\%$). The numerical results of one of the regression coefficients is graphically presented in Figure \ref{fig:result} and shown in Table \ref{tab:mv_by1}. Other regression coefficients performed qualitatively similarly and are available upon request.

	\begin{sidewaystable}[!htbp]
		\centering
		\caption{True and estimated regression coefficients including the 95\% confidence interval for the three imputation strategies under different correlation structures.} 
		\label{tab:mv_by1}
		\begin{tabular}{l
				r@{\hspace{25pt}}rrr@{\hspace{25pt}}rrr@{\hspace{25pt}}rrr@{\hspace{25pt}}rrr}
			\hline
			&& \multicolumn{6}{l}{Monotone} & \multicolumn{6}{l}{Non-monotone}\\ \hline
			& &\multicolumn{3}{l}{Coefficients}  
			& \multicolumn{3}{l}{Coverage} &
			\multicolumn{3}{l}{Coefficients} &  \multicolumn{3}{l}{Coverage} \\ \hline                               
			Scenario & Truth & RI & NI & AI & RI & NI & AI & RI & NI & AI & RI & NI & AI \\  \hline
			1. & 0.08 & 0.08 & 0.08 & 0.08 &   94.3 & 94.7 & 94.8 & 
			0.08 & 0.06 & 0.06 &  94.3 & 97.8 & 97.6 \\ 
			2. & 0.23 & 0.23 & 0.23 & 0.23 &   94.1 & 94.3 & 93.7 & 
			0.23 & 0.17 & 0.17 &  94.0 & 90.3 & 90.2 \\ 
			3. & 0.42 & 0.42 & 0.42 & 0.42 &  95.0 & 95.7 & 95.2 & 
			0.41 & 0.28 & 0.28 &  94.8 & 47.1 & 54.4 \\ 
			4. & 1.17 & 1.17 & 1.17 & 1.17 &   94.7 & 95.0 & 94.8 & 
			1.17 & 0.58 & 0.58 &   95.0 & 1.4 & 3.4 \\ 
			5. & 0.07 & 0.07 & 0.07 & 0.07 &   94.5 & 94.5 & 94.1 &
			0.07 & 0.05 & 0.05 &   94.9 & 97.8 & 97.5 \\ 
			6. & 0.19 & 0.19 & 0.19 & 0.19 &   94.6 & 95.3 & 94.8 &
			0.19 & 0.14 & 0.14 & 94.8 & 93.4 & 93.5 \\ 
			7. & 0.31 & 0.31 & 0.31 & 0.31 &  94.5 & 95.2 & 94.6 & 
			0.31 & 0.22 & 0.22 &   94.8 & 73.7 & 74.7 \\ 
			8. & 0.57 & 0.57 & 0.57 & 0.57 &   94.0 & 95.0 & 95.0 & 
			0.57 & 0.29 & 0.29 & 95.2 & 4.4 & 9.9 \\ 
			9. & 0.06 & 0.06 & 0.06 & 0.06 &   94.1 & 96.0 & 95.5 & 
			0.06 & 0.05 & 0.05 &   95.0 & 98.5 & 98.1 \\ 
			10. & 0.16 & 0.16 & 0.16 & 0.16 &   94.3 & 95.0 & 95.1 & 
			0.16 & 0.12 & 0.12 &   93.3 & 95.8 & 95.5 \\ 
			11. & 0.25 & 0.25 & 0.25 & 0.25 &  94.5 & 94.3 & 93.8 & 
			0.25 & 0.18 & 0.18 & 94.7 & 87.0 & 86.5 \\ 
			12. & 0.40 & 0.40 & 0.40 & 0.40 &  95.0 & 95.5 & 95.2 & 
			0.40 & 0.26 & 0.26 & 94.3 & 41.4 & 47.2 \\ 
			13. & 0.05 & 0.05 & 0.05 & 0.05 &  94.2 & 94.3 & 93.8 & 
			0.06 & 0.04 & 0.04 &  94.8 & 99.2 & 97.8 \\ 
			14. & 0.14 & 0.14 & 0.14 & 0.14 & 95.0 & 94.9 & 94.8 & 
			0.14 & 0.10 & 0.10 & 94.5 & 97.1 & 97.0 \\ 
			15. & 0.21 & 0.21 & 0.21 & 0.21 &  95.3 & 96.2 & 95.0 & 
			0.21 & 0.14 & 0.14 & 94.8 & 93.4 & 93.3 \\ 
			16. & 0.35 & 0.35 & 0.35 & 0.35 &   95.3 & 96.3 & 95.0 & 
			0.35 & 0.25 & 0.25 & 95.1 & 78.2 & 80.0 \\ 
			\hline
			\multicolumn{14}{l}{\textit{Note.} RI = re-imputation, NI = nested imputation, AI = appended imputation.}\\  \hline
		\end{tabular}
	\end{sidewaystable}
	
	\subsubsection{Efficiency}
	The 95\% confidence intervals of variable means had similar relative widths (nested imputation vs. re-imputation: $0.97-1.18$; nested imputation vs. appended imputation: $0.91-1.01$; appended imputation vs. re-imputation: $0.98-1.20$). Estimation of egression coefficients was slightly more efficient under nested imputation compared to re-imputation ($0.87-0.93$) or appended imputation ($0.88-0.92$).  
	
	\subsubsection{Summary}
	When taken together, we found that:
	\begin{enumerate}
	\item All techniques performed similarly under monotone missingness ;
	\item Appended and nested imputation were on par on all scenarios;
	\item Appended and nested imputation performed well when correlations within time points were high;
	\item Appended and nested imputation performed quite bad when correlations between timepoints were high and correlations within timepoints were low.
	\end{enumerate}
	
	\section{Data application}\label{sec:application}
	In practice, violating congeniality in nested and appended imputation may have less severe consequences than our simulation suggested. Real datasets often contain auxiliary variables that provide extra information to reduce the impact of ignoring future data \citep{daniels2014,Xie2015}. To compare the performances of the imputation strategies in practice, we applied them to the POPS dataset \citep{verloove1986}. 
		
	\subsection{Method}
	We predicted five outcomes at age 19 ($t_2$ data: length, cognition, health-related quality of life, internalizing problems, and externalizing problems) from four types of catch-up growth ($t_1$ data: weight, length, head circumference, and weight-length), either unadjusted or adjusted for potential confounders (also $t_1$ data) using analysis models specified by Van Dommelen et al. \citep{dommelen2014}. Potential confounders were neonatal factors (gestational age, and sex) and environmental factors (maternal age at birth, maternal smoking during pregnancy, maternal diabetes, socioeconomic status, parity, ethnicity, and target length). The researchers selected $n=334$ cases born small for gestational age without severe complications ($n=228$ for weight, $n=203$ for length, $n=178$ for head circumference, and $n=64$ for weight adjusted for length) from the incomplete POPS cohort.	A more elaborate descriptions of case selection and operationalization of variables can be found in the original article \citep{dommelen2014}.
		
	We completed the dataset using the three imputation strategies. Similar to the original article, we re-imputed the incomplete $t_1$ and $t_2$ data $m=10$ times \citep{dommelen2014} and we imputed $t_1$ data $m_1=10$ times under nested and appended imputation. We added the incomplete $t_2$ data to each completed $t_1$ dataset and imputed these partially completed datasets $m_2=10$ (nested imputation) and $m_2=1$ (appended imputation) times. 
		
	After multiple imputation, we made the aforementioned data selections and fitted eight linear regression models per outcome variable to the appropriate data. The models predicted the outcome from catch-up growth in weight, length, head circumference or weight adjusted for length, either unadjusted or adjusted for potential confounders. Quantities of interest were regression coefficients of catch-up growth predictors and their $95\%$ confidence intervals.
		
	We performed statistical analyses using \texttt{R} \citep{RCT2016}. 
	We imputed data with \texttt {mice} \citep{buuren2011} 
	using predictive mean matching similar to the original study. 
		
	\subsection{Results}
	\subsubsection{Assumptions for congeniality}
	
	Although none of the data selections followed a strictly monotone missingness pattern, at least $60\%$ of missing values came from cases without observations in $t_2$ data, except for the weight-length predictor ($45.1-64.7\%$ monotone). 

	Potential problems arising from nonmonotone missingness may be mitigated by strong correlations within waves. Each catch-up growth predictor had at least one correlation with another catch-up growth predictor (i.e. within $t_1$) that exceeded the correlation with each of the outcomes (i.e. between $t_1$ and $t_2$). Hence, we considered nested and appended imputation appropriate for these data.
	
	\subsubsection{Parameter estimates}
	Regression coefficients of catch-up growth predicting length at age 19 are presented in Table \ref{tab:application}. The three imputation strategies resulted in similar point estimates and confidence intervals, and agreed on their conclusions in seven out of eight models (weight, length, head circumference; either adjusted or unadjusted, and weight-length unadjusted). The adjusted model of weight-length resulted in contrasting conclusions: Weight-length predicted length after re-imputation (CI: $-0.76- -0.04$), but not after nested and appended imputation (CI nested: $-0.73 - 0.00$; CI appended: $-0.72 - 0.05$). Over all models, nested imputation and re-imputation were approximately equally efficient (relative width 95\% CI nested vs. re-imputation:  $0.85-1.07$) and more efficient than appended imputation (nested: $0.74-0.97$; re-imputation: $0.76-0.95$). Results of other outcome variables were qualitatively similar and are available upon request.
	
	\begin{table}[ht]
		\centering
		\caption{Regression coefficients and (the width of) their 95\% confidence intervals of early catch-up growth predicting length at age 19, after re-imputation, nested imputation and appended imputation. Regression coefficients were unadjusted and adjusted for potential confounders.}
		\label{tab:application}
		\resizebox{\textwidth}{!}{%
			\begin{tabular}{lrR{25pt}@{\hskip3pt} R{5pt} @{\hspace{-2pt}} R{25pt}rp{0.02\textwidth}rR{25pt}@{\hskip3pt} R{5pt} @{\hspace{-2pt}}R{25pt}rp{0.02\textwidth}rR{25pt}@{\hskip3pt} R{5pt} @{\hspace{-2pt}}R{25pt}r}
				\hline
				\multicolumn{18}{l}{Unadjusted}   \\ \hline 
				& \multicolumn{6}{l}{Re-imputation} & \multicolumn{6}{l}{Nested imputation} & \multicolumn{5}{l}{Appended imputation}\\

				& \multicolumn{1}{l}{b} & \multicolumn{3}{c}{95\% CI} & Width && \multicolumn{1}{l}{b} & \multicolumn{3}{c}{95\% CI} & Width & &
				\multicolumn{1}{l}{b} & \multicolumn{3}{c}{95\% CI} & Width \\ \hline  
				$\text{Weight}$ & 0.47 & 0.30 &-& 0.63&  0.33 && 0.46 & 0.29 &-& 0.62 & 0.33 && 0.47 & 0.28 &-& 0.66 &  0.38 \\ 
				$\text{Length}$ & 0.58 & 0.46 &-& 0.71 & 0.25 && 0.57  & 0.44 &-& 0.69 & 0.25 && 0.56 & 0.42 &-& 0.70 & 0.27 \\ 
				$\text{HC}$ & 0.29 & 0.12 &-& 0.45 & 0.33& & 0.26 & 0.10 &-& 0.42 & 0.32 && 0.26 & 0.05 &-& 0.46 & 0.41 \\ 
				$\text{WL}$ & -0.45 & -0.81 &-& -0.10 & 0.71 && -0.41 & -0.77 &-& -0.06 & 0.72 && -0.41& -0.78 &-& -0.04 & 0.74 \\ \hline
				\multicolumn{18}{l}{Adjusted} \\ \hline 
				& \multicolumn{6}{l}{Re-imputation} & \multicolumn{6}{l}{Nested imputation} & \multicolumn{5}{l}{Appended imputation}\\			
				& \multicolumn{1}{l}{b} & \multicolumn{3}{c}{95\% CI} & Width && \multicolumn{1}{l}{b} & \multicolumn{3}{c}{95\% CI} & Width && 
				\multicolumn{1}{l}{b} & \multicolumn{3}{c}{95\% CI} & Width \\  
				\hline
				$\text{Weight}$ & 0.27 & 0.13 &-& 0.42 & 0.29 && 0.26 & 0.10 &-& 0.41 & 0.31& & 0.27 & 0.08 &-& 0.46 & 0.38 \\ 
				$\text{Length}$ & 0.42 & 0.31 &-& 0.53 & 0.23 && 0.40 & 0.28 &-& 0.52 & 0.24 && 0.39 & 0.24 &-& 0.54 & 0.30 \\ 
				$\text{HC}$ & 0.12 & -0.05 &-& 0.29 & 0.34 && 0.09 & -0.06 &-& 0.23 & 0.29 && 0.09 & -0.10 &-& 0.29 & 0.39 \\ 
				$\text{WL}$ & -0.40 & -0.76 &-& -0.04 & 0.73 && -0.36 & -0.73 &-& 0.00 & 0.73 && -0.33 & -0.72 &-& 0.05 & 0.78 \\ \hline
				\multicolumn{18}{l}{\raggedright \textit{Note.} HC = head circumference, WL = Weight adjusted for length.}
			\end{tabular}
		}
	\end{table} 
	
	\section{Discussion}\label{sec:discussion}
	The current study investigated appended imputation as an alternative to re-imputation and nested imputation as generic strategies to deal with multistage incomplete data. Appended imputation is attractive since 1.) it keeps the old imputations in place, thereby preserving the validity of statistical analyses performed on the earlier waves; and 2.) it is computationally and logistically more convenient than nested imputation. We investigated the inherent dangers of appended imputation, and found that it could used for missing data patterns close to monotone, and for situations where the correlations relating variables within waves dominate the correlations relating variables between waves. Especially longtudinal datasets suffering from dropout could benefit from appended imputation. We do not recommend appended imputation over re-imputation when correlations between timepoints are high and correlations within timepoints are low, unless there is an explicit desire to maintain reproducibility of historic at the expense of statistical validity.
	\bigskip
		
	\newpage
	\bibliographystyle{plainnat}
	\bibliography{bibliography}

\begin{thebibliography}{18}
\providecommand{\natexlab}[1]{#1}
\providecommand{\url}[1]{\texttt{#1}}
\expandafter\ifx\csname urlstyle\endcsname\relax
  \providecommand{\doi}[1]{doi: #1}\else
  \providecommand{\doi}{doi: \begingroup \urlstyle{rm}\Url}\fi

\bibitem[Daniels et~al.(2014)Daniels, Wang, and Marcus]{daniels2014}
MJ~Daniels, Chenguang Wang, and BH~Marcus.
\newblock Fully bayesian inference under ignorable missingness in the presence
  of auxiliary covariates.
\newblock \emph{Biometrics}, 70\penalty0 (1):\penalty0 62--72, 2014.

\bibitem[Gelman and Raghunathan(2001)]{gelman2001}
Andrew Gelman and Trivellore~E Raghunathan.
\newblock Using conditional distributions for missing-data imputation.
\newblock \emph{Statistical Science}, 15:\penalty0 268--69, 2001.

\bibitem[Harel(2003)]{harel2003a}
O.~Harel.
\newblock \emph{Strategies for data analysis with two types of missing values}.
\newblock PhD thesis, Citeseer, 2003.

\bibitem[Harel and Schafer(2003)]{harel2003b}
O.~Harel and J.~Schafer.
\newblock Multiple imputation in two stages.
\newblock In \emph{Proceedings of Federal Committee on Statistical Methodology
  2003 Conference}. Citeseer, 2003.

\bibitem[Hille et~al.(2005)Hille, Elbertse, Gravenhorst, Brand,
  Verloove-Vanhorick, et~al.]{hille2005}
ETM Hille, L~Elbertse, J~Bennebroek Gravenhorst, Ren{\'e} Brand,
  SP~Verloove-Vanhorick, et~al.
\newblock Nonresponse bias in a follow-up study of 19-year-old adolescents born
  as preterm infants.
\newblock \emph{Pediatrics}, 116\penalty0 (5):\penalty0 e662--e666, 2005.

\bibitem[McGinniss and Harel(2016)]{mcginniss2016}
J.~McGinniss and O.~Harel.
\newblock Multiple imputation in three or more stages.
\newblock \emph{Journal of Statistical Planning and Inference}, 176:\penalty0
  33--51, 2016.

\bibitem[Meng(1994)]{meng1994}
Xiao-Li Meng.
\newblock Multiple-imputation inferences with uncongenial sources of input.
\newblock \emph{Statistical Science}, pages 538--558, 1994.

\bibitem[{R Core Team}(2016)]{RCT2016}
{R Core Team}.
\newblock \emph{R: A Language and Environment for Statistical Computing}.
\newblock R Foundation for Statistical Computing, Vienna, Austria, 2016.
\newblock URL \url{http://www.R-project.org/}.

\bibitem[Rubin(1987)]{rubin1987}
D.~B. Rubin.
\newblock \emph{Multiple imputation for nonresponse in surveys}, volume~81.
\newblock John Wiley \& Sons, 1987.

\bibitem[Rubin(2003)]{rubin2003}
D.~B. Rubin.
\newblock Nested multiple imputation of {NMES} via partially incompatible
  {MCMC}.
\newblock \emph{Statistica Neerlandica}, 57\penalty0 (1):\penalty0 3--18, feb
  2003.

\bibitem[Rubin(1996)]{rubin1996}
Donald~B Rubin.
\newblock Multiple imputation after 18+ years.
\newblock \emph{Journal of the American statistical Association}, 91\penalty0
  (434):\penalty0 473--489, 1996.

\bibitem[Shen(2000)]{shen2000}
Zijin Shen.
\newblock \emph{Nested multiple imputation}.
\newblock PhD thesis, Harvard University, Cambridge, MA., 2000.

\bibitem[Van~Buuren and Groothuis-Oudshoorn(2011)]{buuren2011}
S.~Van~Buuren and K.~Groothuis-Oudshoorn.
\newblock \texttt{mice}: Multivariate imputation by chained equations in r.
\newblock \emph{Journal of Statistical Software}, 45\penalty0 (3), 2011.

\bibitem[Van~Dommelen et~al.(2014)Van~Dommelen, Van Der~Pal,
  Bennebroek~Gravenhorst, Walther, Wit, and Van der Pal~de Bruin]{dommelen2014}
Paula Van~Dommelen, Sylvia~M Van Der~Pal, J~Bennebroek~Gravenhorst, Frans~J
  Walther, Jan~M Wit, and KM~Van der Pal~de Bruin.
\newblock The effect of early catch-up growth on health and well-being in young
  adults.
\newblock \emph{Annals of Nutrition and Metabolism}, 65\penalty0
  (2-3):\penalty0 220--226, 2014.

\bibitem[Veen et~al.(1991)Veen, Ens-Dokkum, Schreuder, Verloove-Vanhorick,
  Ruys, and Brand]{veen1991}
Sylvia Veen, Martina~H Ens-Dokkum, Anneke~M Schreuder, S~Pauline
  Verloove-Vanhorick, JH~Ruys, and R~Brand.
\newblock Impairments, disabilities, and handicaps of very preterm and
  very-low-birthweight infants at five years of age.
\newblock \emph{The Lancet}, 338\penalty0 (8758):\penalty0 33--36, 1991.

\bibitem[Verloove-Vanhorick et~al.(1986)Verloove-Vanhorick, Verwey, Brand,
  Gravenhorst, Keirse, and Ruys]{verloove1986}
S~Pauline Verloove-Vanhorick, RA~Verwey, R~Brand, J~Bennebroek Gravenhorst,
  MJNC Keirse, and JH~Ruys.
\newblock Neonatal mortality risk in relation to gestational age and
  birthweight: results of a national survey of preterm and very-low-birthweight
  infants in the netherlands.
\newblock \emph{The Lancet}, 327\penalty0 (8472):\penalty0 55--57, 1986.

\bibitem[Xie and Meng(2015)]{Xie2015}
Xianchao Xie and Xiao-Li Meng.
\newblock Dissecting multiple imputation from a multi-phase inference
  perspective: What happens when god's, imputer's and analyst's models are
  uncongenial.
\newblock \emph{Statist. Sinica}, 2015.

\bibitem[Zhu and Raghunathan(2015)]{zhu2015}
Jian Zhu and Trivellore~E Raghunathan.
\newblock Convergence properties of a sequential regression multiple imputation
  algorithm.
\newblock \emph{Journal of the American Statistical Association}, 110\penalty0
  (511):\penalty0 1112--1124, 2015.

\end{thebibliography}
	
\end{document}